\documentclass[12pt,a4paper]{article}

\usepackage{amsmath}
\usepackage{amssymb}
\usepackage{caption}
\usepackage{subcaption}
\usepackage{indent first}
\usepackage{jheppub}
\usepackage{xspace}

\newcommand{\bZ}{\mathbb{Z}}
\def\Nequals#1{$\mathcal{N}{=}#1$}
\def\SU{\mathrm{SU}}
\def\Sp{\mathrm{Sp}}

\def\U{\mathrm{U}}
\def\SO{\mathrm{SO}}
\def\tr{\mathop{\mathrm{tr}}\nolimits}

\def\Tr{\mathop{\mathrm{Tr}}\nolimits}

\def\vev#1{\langle#1\rangle}

\begin{document}

\title{Anomaly of strings of 6d \Nequals{(1,0)} theories}
\abstract{
We obtain the anomaly polynomial of strings of general 6d \Nequals{(1,0)} theories in terms of  anomaly inflow. 
Our computation sheds some light on the reason why the simplest 6d \Nequals{(1,0)} theory has $E_8$ flavor symmetry, and also partially explains a curious numerology in F-theory.
}
\author{Hiroyuki Shimizu}
\author{and Yuji Tachikawa}
\affiliation{Kavli Institute for the Physics and Mathematics of the Universe, \\
the  University of Tokyo,  Kashiwa, Chiba 277-8583, Japan}
\preprint{IPMU16-0121}

\maketitle

\section{Introduction}\label{sec:introduction}
We have seen a renaissance of the study of 6d \Nequals{(1,0)} theories since the publication of the seminal work \cite{Heckman:2013pva}.
For example, we now have a better understanding of the landscape of such theories \cite{DelZotto:2014hpa,Heckman:2015bfa,Bhardwaj:2015xxa,Cordova:2015fha,Heckman:2015axa} and have a general formula for their anomaly polynomials \cite{Ohmori:2014kda,Intriligator:2014eaa}.

An essential feature of these theories is that they have supersymmetric strings charged under self-dual tensor fields. The properties of these strings have received much attention, but the analysis has so far been mostly restricted to the case when the description in terms of the refined topological string and/or the 2d gauge theory is available \cite{Haghighat:2013gba,Haghighat:2013tka,Haghighat:2014pva,Kim:2014dza,Haghighat:2014vxa,Hohenegger:2015cba,Gadde:2015tra,Kim:2015fxa,Haghighat:2016jjf}.

In this note, we provide a formula for the anomaly polynomial of the 2d worldsheet theories of these strings. The only input required is  the anomaly of the bulk 6d theory, and therefore our formula applies generically. Our formula is given in \eqref{eq:generalmainformula}, \eqref{eq:concretemainformula}.

When a 2d gauge theory description of the worldsheet theory of the strings is known, we can compare the outcome of our main formula \eqref{eq:generalmainformula}, \eqref{eq:concretemainformula} and the anomaly polynomial  computed from the gauge theory spectrum. We will see below that these two computations indeed do match.
When such a 2d gauge theory description is not known, our formula should help us look for one.

Prime examples where the 2d gauge theories are not known are the following. 
Take an F-theory compactification on a complex two-dimensional base with an isolated $\mathbb{P}^1$ of self intersection $-n$ with $n\ge 3$. 
It is known that this configuration automatically forces a nontrivial gauge algebra $\mathfrak{g}$ on the curve.

\begin{table}
\[
\begin{array}{l|ccccccccc}
n & 3 & 4 & 5 & 6 & 8 & 12\\
\hline
\mathfrak{g} & \mathfrak{su}(3) & \mathfrak{so}(8) & \mathfrak{f}_4 & \mathfrak{e}_6 & \mathfrak{e}_7 & \mathfrak{e}_8 \\
\hline
h^\vee & 3 & 6 & 9 & 12 & 18 & 30
\end{array}
\]
\caption{Smallest allowed gauge algebra $\mathfrak{g}$ and its dual Coxeter number $h^\vee$ on an isolated curve of self-intersection $-n$ in F-theory. We see a relation $h^\vee=3(n-2).$\label{table:foo}}
\end{table}

In Table~\ref{table:foo} we listed this $\mathfrak{g}$ together with its dual Coxeter number $h^\vee$ for the cases with no additional matter on the curve; for a derivation, see e.g.~\cite{Morrison:2012np}.%
\footnote{\label{footnote:deligne}As an aside, let us point out curious coincidences concerning Table~\ref{table:foo}: this is an exact subset of Table~4 of \cite{Beem:2013sza} and Table~4 of \cite{Lemos:2015orc}, where a different topic, namely the 2d chiral algebras associated to 4d \Nequals2 theories were studied. The two tables in \cite{Beem:2013sza,Lemos:2015orc} contain $\mathfrak{su}(2)$ and $\mathfrak{g}_2$ in addition, and $k_{4d}$ there equals $n$ here. This series of groups with an addition of the empty group,  namely \[
\varnothing, \mathfrak{su}(2) , \mathfrak{su}(3) , \mathfrak{g}_2 , \mathfrak{so}(8) , \mathfrak{f}_4 , \mathfrak{e}_6 , \mathfrak{e}_7 , \mathfrak{e}_8,
\] is known as Deligne's exceptional series of groups, from the papers by Deligne \cite{Deligne1,Deligne2} in 1996.  The relation of the F-theory list of groups and Deligne's exceptional series was already noted in a paper by Grassi and Morrison \cite{Grassi:2000we}, see Lemma 7.4 there.

It is to be noted that the same series of groups has also appeared in a paper by Mathur, Mukhi and Sen \cite{Mathur:1988na} from 1988 when 2d RCFTs with only two characters were systematically looked for. 
Is there a single unifying principle behind these appearance of the same sequence of groups in various corners of string theory?
(Deligne's exceptional series of groups was also independently found by Cvitanovi\'c, see Sec.~21.2 of \cite{Cvitanovic} for the history of multiple independent (re)discoveries of this series.
The authors thank L. Rastelli for the information.)
} 
This configuration gives rise to a 6d \Nequals{(1,0)} system with one tensor multiplet and a vector multiplet for $\mathfrak{g}$. These 6d theories are known as minimal 6d \Nequals{(1,0)} theories.

Instantons of this vector multiplet become supersymmetric strings with \Nequals{(0,4)} supersymmetry. 
For $n=4$ there is a natural gauge theory description and detailed studies have been made \cite{Haghighat:2014vxa,Gadde:2015tra} but not much is known in other cases.%
\footnote{But see Note Added below.}
Our formula~\eqref{eq:generalmainformula}, \eqref{eq:concretemainformula} gives at least the anomaly polynomials of these strings,
and also explains  a curious numerology \begin{equation}
h^\vee = 3(n-2)\label{eq:curious}
\end{equation} 
that is evident in Table~\ref{table:foo}, as we will see below.

In the rest of the note, we first provide the derivation of the formula \eqref{eq:generalmainformula}, \eqref{eq:concretemainformula} using the anomaly inflow in Sec.~\ref{sec:derivation}. We also provide checks by comparing with the anomaly polynomials computed from 2d gauge theories when available.
Then we discuss some of the  implications of our formula in Sec.~\ref{sec:implications} and conclude.

\paragraph{Note Added:} In the final stage of the research leading to this note, a paper \cite{Kim:2016foj} appeared, in which the authors also presented  our main formula \eqref{eq:generalmainformula}, although their main concern was the elliptic genera of the strings.
In \cite{Kim:2016foj} a 2d gauge theory description was also given for the case $n=3$.

\section{Derivation and Checks}\label{sec:derivation}

\subsection{Anomaly formula from the inflow}
\paragraph{Review of the tensor branch effective action.}
Let us briefly recall the tensor branch effective action of 6d \Nequals{(1,0)} theories to set up the notations. On the generic point of the tensor branch, the 6d \Nequals{(1,0)} theory consists of massless tensor/vector/hyper multiplets plus massive dynamical stringy excitations. Integrating out the massive modes, a part of the effective action is given as
\begin{equation}
2 \pi  \int  \eta^{ij} \Bigl(\frac{1}{2}dB_i \wedge \star dB_j 
+ 
  B_i  \wedge I_j \Bigr).\label{eq:bulkaction}
\end{equation}
Here $B_i$ is the self-dual 2-form field normalized so that its field strength is quantized in integer values.\footnote{Because $B_i$ is self-dual, it is imprecise to write the kinetic term as in \eqref{eq:bulkaction}, but we will see that it is convenient to include it here for the inflow computation.
} The strings in the 6d theory are charged under those 2-form fields and the charge matrix is given by $\eta^{ij}$. The Dirac quantization law requires that the matrix $\eta^{ij}$ is symmetric, positive definite and integral. In the following, indices are raised/lowered by using $\eta^{ij}$. 

The Green-Schwarz coupling $B_i \wedge I_j$ contributes to the anomaly polynomial of the 6d theory by $\frac12 \eta^{ij}I_i \wedge I_j$ \cite{Sagnotti:1992qw}.  For all  6d \Nequals{(1,0)} theories known to exist so far, the 4-form $I_i$ is known to have a concrete form given by  
\begin{equation}
\eta^{ij} I_j = \frac14 \biggl{(}\eta^{ia}\Tr F_a^2 - (2-\eta^{ii}) p_1(T)\biggr{)} + h^{\vee}_{G_i} c_2(I)\label{eq:GSterm}
\end{equation}
as derived in \cite{Sadov:1996zm, Ohmori:2014kda}.
Here the sum over the indices $j$ and $a$ is taken, but the index in $\eta^{ii}$ is not summed. The field strengths $F_a$ include both dynamical and background gauge fields. Accordingly, we extend the charge matrix $\eta$ to include both dynamical and background tensor multiplets.\footnote{The indices $a,b, \dots$ are for both dynamical and background fields, while the indices $i,j,\dots$ are only for dynamical ones.} 
$p_1(T)$ is the first Pontrjagin class of the tangent bundle of the 6d spacetime and 
$c_2(I)$ is the second Chern class of the background $\SU(2)_I$ R-symmetry bundle of the 6d \Nequals{(1,0)} supersymmetry.
The trace $\Tr$ is normalized so that $\frac14 \Tr F^2$ is $1$ for one-instanton configurations. $h^\vee_{G_i}$ is the dual Coxeter number of the Lie group $G_i$. When $G_i=\varnothing$, it is understood as $1$, which happens only when $\eta^{ii}=1$ or $2$.

\paragraph{Main formula.} We put the  self-dual string at $x_2 = \cdots = x_5=0$ in 6d spacetime. The charge of the string is specified by a vector $Q_i$ with integral entries. The worldvolume theory on the string has the global symmetries $\SU(2)_L \times \SU(2)_R \times \SU(2)_I \times \prod_a G_a$. Here $\SU(2)_L \times \SU(2)_R\simeq \SO(4)_N$ comes from rotating the normal directions to the string, while $\SU(2)_I \times \prod_a G_a$ comes from the R, gauge and global symmetries of the bulk 6d theory.\footnote{Here we should not confuse $\SU(2)_R$ with the R-symmetry of 6d supersymmetry.}

The supercharge of 6d \Nequals{(1,0)} theory decomposes as \begin{equation}
(\mathbf{2},\mathbf{1},\mathbf{2})_+
+(\mathbf{1},\mathbf{2},\mathbf{2})_-
\end{equation} under $\SU(2)_L\times\SU(2)_R\times \SU(2)_I$, with the subscript $\pm$ denoting the chirality. 
We take the convention that the remaining half of the supercharge is in $(\mathbf{1},\mathbf{2},\mathbf{2})_-$.
Therefore, the $\SO(4)$ R-symmetry of \Nequals{(0,4)} supersymmetry is identified with $\SU(2)_R\times \SU(2)_I$.

Then, the anomaly 4-form $I_4$ of the 2d \Nequals{(0,4)} theory on the string is given as
\begin{equation}
I_4 = \frac{\eta^{ij}Q_i Q_j}{2} \biggl{(}c_2(L)-c_2(R)\biggr{)} 
+\eta^{ij} Q_i I_j \label{eq:generalmainformula}.
\end{equation}

If we assume the general validity of the concrete formula \eqref{eq:GSterm}, this can be more explicitly written as \begin{multline}
I_4 = \frac{\eta^{ij}Q_i Q_j}{2} \biggl{(}c_2(L)-c_2(R)\biggr{)} \\
+ Q_i \biggl{(} \frac14 \eta^{ia} \Tr F^2_a - \frac{2-\eta^{ii}}{4}(p_1(T) - 2 c_2(L) -2c_2(R)) + h^\vee_{G_i} c_2(I) \biggr{)}  \label{eq:concretemainformula}.
\end{multline}
In deriving \eqref{eq:generalmainformula} and \eqref{eq:concretemainformula}, we decompose the 6d $p_1(T)$ as 2d $p_1(T)+p_1(N)$ and we use the relations $\chi(N) = c_2(L) - c_2(R)$ and $p_1(N) = -2 c_2(L)-2c_2(R)$. 

In the following, we will derive the formula \eqref{eq:generalmainformula} using the inflow computation. We will then check \eqref{eq:concretemainformula} against concrete examples for which 2d gauge theory description is known.

\paragraph{Inflow argument.} Here we perform the anomaly inflow computation for the self-dual string, using a method pioneered by \cite{Freed:1998tg}.
The anomaly inflow of self-dual strings was studied in other places before, see e.g.~\cite{Henningson:2004dh,Berman:2004ew,Henningson:2005hd}, mostly in the case of 6d  \Nequals{(2,0)} theory.

In the presence of the string, the Bianchi identity for the 3-form field strength is modified to be
\begin{equation}
dH_i = I_i + Q_i \prod_{j=2}^{5} \delta (x_j)dx_j, \label{eq:Bianchi}
\end{equation}
where $Q_i$ is the charge of the string. Its solution is given by 
\begin{equation}
H_i = Q_i \frac{e^{(0)}_3}{2} + (\text{regular}),\label{eq:BIanchisol}
\end{equation}
where $e^{(0)}_3$ is the global angular form of the $S^3$ bundle of the tubular neighborhood of the string, which is related to the Euler class $\chi_4(N)$ of the normal bundle by $de^{(0)}_3 = 2\chi_4(N)$.

To compute the anomaly polynomial of the 2d theory, it is convenient to use 
\begin{equation}
2\pi \int_{Y_7} \eta^{ij} \Bigl( \frac{1}{2}dH_i \wedge H_j 
+ H_i  \wedge I_j \Bigr),\label{eq:7-form}
\end{equation}
instead of \eqref{eq:bulkaction}. Here $Y_7$ is an auxiliary 7d manifold bounding the physical 6d spacetime. We also extend  the worldsheet of the string to $Y_7$ and denote it as $M_3$. 

In the presence of the string, the most singular term in \eqref{eq:7-form} is given as
\begin{equation}
2\pi \int_{Y_7} \Bigl( \frac{\eta^{ij}Q_i Q_j}{4} \chi_4(N) e^{(0)}_3 + \eta^{ij}Q_i I_j \frac{e^{(0)}_3}{2} \Bigr).\label{eq:singular}
\end{equation}
Anomaly inflow tells us that integrating out \eqref{eq:singular} we obtain the anomaly on the string worldsheet. Integration is straightforward since there is a factor of the Euler class $\chi_4(N)$ which just reduces the integral over $Y_7$ to $M_3$. The result is
\begin{equation}
2\pi \int_{M_3} \Bigl( \frac{\eta^{ij}Q_i Q_j}{4}  e^{(0)}_3 + \eta^{ij}Q_i I^{(0)}_{j,3} \Bigr)
\end{equation}
which correctly reproduces the formula \eqref{eq:generalmainformula}.

The computation of the contribution from the kinetic term of \eqref{eq:bulkaction} or equivalently the first term of \eqref{eq:7-form}, in the form presented above, involves some  amount of hand-waving, due to the self-dual nature of the tensor fields. 
A more careful derivation, following the one presented in \cite{Kim:2012wc} in the case of D3-branes coupled to a self-dual 5-form, gives the same result.

\subsection{Comparison with gauge theory description}
When the UV realization of the string worldsheet theory as a 2d \Nequals{(0,4)}  gauge theory is known, we can easily compute its anomaly polynomial by counting the number of multiplets.\footnote{A nice summary of multiplets of \Nequals{(0,4)} supersymmetric theories can be found e.g.~in \cite{Tong:2014yna}.
As for the conventions,  a left moving (plus sign in the chirality) complex Weyl fermion in a representation $\rho$ contributes to the anomaly polynomial by $\hat{A}(T) \tr_{\rho} \text{e}^{iF}$. In particular, a left moving complex Weyl fermion in 2d in the fundamental representation of $\SU(2)$ gives the anomaly $-\frac12 \tr_{\text{fund}} F^2 = -\frac14 \Tr F^2 = - c_2(\SU(2))$. The numerical factor relating $\Tr F^2$ and $\tr_{\text{fund}} F^2$ for various groups is summarized e.g.~in the appendix of \cite{Ohmori:2014kda}.} In this subsection, we provide further pieces of evidence for the formula \eqref{eq:concretemainformula} by checking the agreement. 

\paragraph{E-string theory.}
On the one hand, using our formula \eqref{eq:concretemainformula}, we see the anomaly 4-form of the bound state of $Q$ E-strings is given as 
\begin{equation}
I_4^{\text{E-string}}(Q) = \frac{Q^2 + Q}{2} c_2(L) - \frac{Q^2 - Q}{2} c_2(R) - \frac{Q}4 \Tr F^2_{E_8} - \frac{Q}4 p_1(T) + Q c_2(I).\label{eq:Estringresult}
\end{equation}
Here we used the fact that $\eta^{ia}$ in \eqref{eq:GSterm} for the $E_8$ global symmetry is $-1$.

On the other hand, the matter content of the gauge theory was determined in \cite{Kim:2014dza} and is summarized in Table \ref{table:Estring}.
\begin{table}
\begin{center}
\begin{tabular}{cccc}
 & vector $(A_{\mu},\lambda_{+}^{\dot{\alpha} A})$ & hyper $(\phi_{\alpha \dot{\alpha}},\lambda_{-}^{\alpha A})$ & Fermi $(\Psi_{+}^{l})$ \\
\hline
$\mathrm{O}(Q)$ & antisymmetric & symmetric & fund \\
\hline
$\SU(2)_L$ & - & fund & - \\
$\SU(2)_R$ & fund & - & - \\
$\SU(2)_I$ & fund & fund & - \\
$\SO(16)$ & - & - & fund \\
\end{tabular}
\caption{The gauge theory description of the worldsheet theory on $Q$ E-strings. $\alpha,\dot{\alpha}=1,2$ are indices for $\SU(2)_{L,R}$, $A=1,2$ are for $\SU(2)_I$ and $l=1,\dots,16$ are for $\SO(16)$ flavor symmetry. We explicitly write the representations of the fermions in the multiplets. $O(Q)$ is the gauge symmetry while the other symmetries are global. In the computation of anomalies, we should note that all the fermions listed above are real.  \label{table:Estring}}
\end{center}
\end{table}
Noting that all the fermions in the table are real and we have to multiply $1/2$ in the anomaly computation, we can check that the $\SU(2)_{L,R,I}$ anomaly matches with \eqref{eq:Estringresult}. 

The anomaly of the $E_8$ global symmetry also matches under the assumption that the $\SO(16)$ symmetry in the UV enhances to $E_8$ in the IR. In fact, the Fermi multiplet $\Psi^l_{+}$ contributes to the anomaly by $\frac12 Q (-\frac12 \tr_{\text{fund}} F^2_{\SO(16)}) = -\frac{Q}4 \Tr F^2_{\SO(16)}$.

For the gravitational anomaly, the vector multiplet gives $-\frac1{24} (Q^2-Q) p_1(T)$, the hypermultiplet gives $+\frac1{24}(Q^2 + Q) p_1(T)$ and the Fermi multiplet gives $-\frac13 Q p_1(T)$. Summing up those contributions, we reproduce the coefficient of $p_1(T)$ in \eqref{eq:Estringresult}.

\paragraph{$A_1$ \Nequals{(2,0)} theory.}  According to our formula \eqref{eq:concretemainformula}, the anomaly on the charge $Q$ string in $A_1$-type \Nequals{(2,0)} theory (called M-string in \cite{Haghighat:2013gba}) is given as 
\begin{equation}
I^{\text{M-string}}_4 (Q) = Q^2 (c_2(L)-c_2(R)) + Q (c_2(I)-c_2(F)).\label{eq:Mstringresult}
\end{equation}
Here we decompose the $\SO(5)$ R-symmetry as $\SU(2)_F \times \SU(2)_I \subset \SO(5)$ and regard $\SU(2)_I$ as the R-symmetry of 6d \Nequals{(1,0)} supersymmetry and $\SU(2)_F$ as the flavor symmetry. Both $\SU(2)$ symmetries are realized on the worldsheet theory on M-string and appear in the anomaly polynomial \eqref{eq:Mstringresult}.

The gauge theory description of the worldsheet theory \cite{Haghighat:2013gba} is listed in Table \ref{table:Mstring}. It is straightforward to check that the counting of multiplets reproduces \eqref{eq:Mstringresult}.

\begin{table}
\begin{center}
\begin{tabular}{ccccc}
  & vector $(A_{\mu},\lambda_{+}^{\dot{\alpha} A})$ & hyper $(\phi_{\alpha \dot{\alpha}},\lambda_{-}^{\alpha A})$ & hyper $(q_{\dot{\alpha}},\psi_{-}^{A})$ & Fermi $(\psi_{+}^{F})$ \\
\hline
$\mathrm{U}(Q)$ & adjoint & adjoint & fund & fund\\
\hline
$\SU(2)_L$ & - & fund & - & - \\
$\SU(2)_R$ & fund & - & - & -  \\
$\SU(2)_I$ & fund & fund & fund & -\\
$\SU(2)_F$ & - & - & - & fund\\
\end{tabular}
\caption{The  gauge theory on the worldsheet of $Q$ M-strings. Here the indices $\alpha,\dot{\alpha},A$ are the same as in the E-string. $F=1,2$ is the fundamental indices of $\SU(2)_F$ symmetry. $\U(Q)$ is the gauge symmetry and the others are global. Again, the fermions $\lambda$ are real.  \label{table:Mstring}}
\end{center}
\end{table}

\paragraph{$n=4$ minimal 6d \Nequals{(1,0)} theory.} According to \eqref{eq:concretemainformula}, the anomaly 4-form of the charge $Q$ string in $n=4$ minimal 6d \Nequals{(1,0)} theory is given as
\begin{equation}
I_4^{n=4}(Q) =  (2Q^2-Q) c_2(L) -(2Q^2 +Q) c_2(R) + Q \Tr F^2_{\SO(8)} + \frac{Q}2 p_1(T) + 6Q  c_2(I).\label{eq:n=4stringanomaly}
\end{equation}

 The gauge theory on the worldsheet is determined in \cite{Haghighat:2014vxa}. The matter content and representations are given as in Table \ref{table:n=4string}. 
\begin{table}
\begin{center}
\begin{tabular}{cccc}
 & vector $(A_{\mu},\lambda_{+}^{\dot{\alpha} A})$ & hyper $(\phi_{\alpha \dot{\alpha}},\lambda_{-}^{\alpha A})$ & hyper $(q_{\dot{\alpha}},\psi_{-}^{A})$ \\
\hline
$\Sp(Q)$ & symmetric & antisymmetric & fund \\
\hline
$\SU(2)_L$ & - & fund & - \\
$\SU(2)_R$ & fund & - & - \\
$\SU(2)_I$ & fund & fund & fund \\
$\SO(8)$ & - & - & fund \\
\end{tabular}
\caption{The gauge theory on the charge $Q$ string in $n=4$ minimal 6d \Nequals{(1,0)} theory.  $\Sp(Q)$ is the gauge symmetry and $\SO(8), \SU(2)_{L,R,I}$ are global symmetries. The indices $\alpha,\dot{\alpha},A$ are the same as in the E-string case. Again the fermions $\lambda$ are all real. The bifundamental hyper of $\Sp(Q) \times \SO(8)$ is in fact a half-hypermultiplet. \label{table:n=4string}}
\end{center}
\end{table}
We can check that the multiplets in Table \ref{table:n=4string} correctly reproduce the anomaly \eqref{eq:n=4stringanomaly}.

\paragraph{6d string chains.} The paper \cite{Gadde:2015tra} studied the strings in a class of 6d \Nequals{(1,0)} theories engineered by an F-theory compactification on a base with a linear chain of $\mathbb{P}^1$'s with self-intersections $-1,-2,$ and $-4$. The 2d \Nequals{(0,4)} gauge theory on the string worldsheet is a linear quiver gauge theory with gauge groups $\prod_{i=1}^{N}G(Q_i)$. The string charge $\{Q_i\}_{i=1}^{N}$ determines ranks of gauge groups, while precise matter content and types of gauge groups depend on which 6d theory we consider.


For these 2d theories, we can also check the agreement in anomalies computed by the formula \eqref{eq:concretemainformula} and the multiplet counting. As a simple example, let us compute the $\SU(2)_I$ anomaly of the charge-$\{Q_i\}_{i=1}^N$ string in rank-$N$ 6d $(\SU(k),\SU(k))$ conformal matter \cite{DelZotto:2014hpa} in two ways. From the formula \eqref{eq:concretemainformula}, the coefficient of $c_2(I)$ is equal to $k\sum_{i=1}^{N}Q_i$. On the other hand, the multiplets in the quiver description having a non-trivial $\SU(2)_I$ anomaly are $\U(Q_i)$ vectors, $\U(Q_i)$-adjoint hypers and $\U(Q_i)\times \SU(k)_i$-bifundamental hypers \cite{Gadde:2015tra}. However, the contributions from vectors and adjoint hypers cancel out, and the bifundamental hypers have the anomaly $(k\sum_{i=1}^{N}Q_i )c_2(I)$ in total, as expected.

\section{Implications}\label{sec:implications}
\subsection{ADE classification of 6d \Nequals{(2,0)} theories}
Let us start by reviewing a nice argument by Henningson \cite{Henningson:2004dh} for the ADE classification of 6d \Nequals{(2,0)} theories.\footnote{Another field theoretical argument for the ADE classification of 6d \Nequals{(2,0)} theories can be found in \cite{Cordova:2015vwa}.}
Take an \Nequals{(2,0)} theory with $r$ tensor multiplets, on generic points on its tensor branch. 
There will be strings charged under these tensor multiplets.

Given  two strings with charges $\vec Q$ and $\vec Q'$ respectively,  let us write the Dirac pairing as \begin{equation}
\vev{\vec Q,\vec Q'}= \eta^{ij} Q_i Q_j'.
\end{equation}
The Dirac quantization law demands that we have $\vev{\vec Q,\vec Q'}\in \bZ$.

Let us consider a single string with charge $\vec Q$, and let us determine the term proportional to $c_2(L)$, $c_2(R)$, $p_1(T)$ of the anomaly polynomial.
Since the string breaks translational invariance and supertranslational invariance,
there are four bosonic zero-modes and eight chiral Majorana fermionic zero-modes,
forming a hypermultiplet of the worldsheet \Nequals{(4,4)} supersymmetry.
The fermionic zero-modes transform as $(\mathbf{2},\mathbf{1},\mathbf{2})_-
+(\mathbf{1},\mathbf{2},\mathbf{2})_+$
under $\SU(2)_L\times\SU(2)_R\times \SU(2)_I$, with the subscript $\pm$ denoting the chirality. 
In total, the anomaly polynomial is  \begin{equation}
I_4^\text{worldsheet}=(c_2(L)+\frac1{12}p_1(T))-(c_2(R)+\frac1{12}p_1(T))=c_2(L)-c_2(R).
\label{eq:mmm}
\end{equation}

The crucial assumption in \cite{Henningson:2004dh} was that the worldsheet theory is given purely by these Nambu-Goldstone zero modes. 
Then the anomaly \eqref{eq:mmm} needs to be reproduced from the anomaly inflow.  
At this point we do not know the 6d Green-Schwarz coupling. We just assume a generic one \begin{equation}
dH_i=c_i p_1(T)\label{eq:2,0GS}
\end{equation} where we neglected contributions from the 6d R-symmetry $c_2(I)$ since they do not affect the 2d anomaly terms we are considering here.

Using our inflow formula \eqref{eq:generalmainformula}, we obtain \begin{equation}
I_4^\text{worldsheet}=\frac{\vev{\vec Q,\vec Q}}2 (c_2(L)-c_2(R))
+ \vev{\vec Q,\vec c} (p_1(T)-2c_2(L)-2c_2(R)).\label{eq:nnn}
\end{equation}
Comparing \eqref{eq:mmm} and \eqref{eq:nnn}, we find that we need \begin{equation}
\vev{\vec Q,\vec Q}=2,
\end{equation}  and at the same time we conclude $c_i=0$ in \eqref{eq:2,0GS}.

We see that the charge lattice of strings of a 6d \Nequals{(2,0)} theory is an integral lattice generated by vectors whose  length squared is two. 
This condition is known to be equivalent to the fact that the charge lattice is a simply-laced root lattice, and therefore it has an ADE classification.
We also derived $c_i=0$ in \eqref{eq:2,0GS}, which agrees with a different computation done in \cite{Ohmori:2014kda}.

\subsection{Existence of $E_8$ flavor symmetry of the smallest 6d \Nequals{(1,0)} theory}
Let us now proceed to \Nequals{(1,0)} theories. 
Nobody would disagree in saying that the E-string theory is the smallest nontrivial \Nequals{(1,0)} theories in six dimensions. 
Still, this smallest theory somehow has $E_8$ flavor symmetry. Why is that?

Let us try to mimic the argument recalled in the previous subsection. 
Again, we restrict attention to the terms proportional to $c_2(L)$, $c_2(R)$ and $p_1(T)$ in the anomaly polynomial. 

Suppose we have an \Nequals{(1,0)} theory with one tensor multiplet.
To make the 6d theory as small as possible, let us assume that the Dirac pairing is given by
\begin{equation}
\eta=1.\label{eq:eta}
\end{equation}  and there is no dynamical gauge field on the tensor branch. 
As for the Green-Schwarz term, we assume the validity of the general formula \begin{equation}
dH=I=\frac{\eta-2}{4} p_1(T).\label{eq:GS}
\end{equation}  For 6d \Nequals{(1,0)} theories constructed from F-theory, this relation follows from the intersection number of the canonical divisor and the genus-0 curve producing the tensor multiplet \cite{Sadov:1996zm}.
A purely field-theoretical derivation of \eqref{eq:GS} is not known yet to the authors' knowledge, but it should not be impossible to find one.

For a string of charge $Q=1$, by plugging in \eqref{eq:eta} and \eqref{eq:GS} to our formula \eqref{eq:generalmainformula}, we find the inflow  \begin{align}
I_4^{\text{inflow}} &= \frac12 (c_2(L)-c_2(R)) - \frac14 (p_1(T) - 2c_2(L)-2c_2(R)) + c_2(I)\nonumber \\
&= c_2 (L) + c_2(I) - \frac14 p_1(T). \label{eq:inflow}
\end{align} 

On the string worldsheet, there are bosonic and fermionic zero modes coming from the breaking of the bulk translational and supertranslational symmetry. 
They form a hypermultiplet of 2d \Nequals{(0,4)} supersymmetry, and have the anomaly polynomial \begin{equation}
I_4^{\text{zero modes}}=c_2 (L) + c_2 (I) + \frac1{12} p_1(T).\label{eq:goldstone}
\end{equation}

Comparing \eqref{eq:inflow} and \eqref{eq:goldstone}, we know that there necessarily are some additional degrees of freedom on the worldsheet since there is a mismatch in the gravitational anomaly by $-\frac13 p_1(T)$. To account for the difference, the simplest possibility is to add a chiral CFT on the non-supersymmetric side of the string worldsheet, with $c=8$.

Assuming that the partition function of the string worldsheet theory is well-defined up to a phase, this additional chiral CFT with $c=8$ needs to be a theory with only a single character. 
This forces us to choose the $E_8$ current algebra of level one.
We cannot say that the argument above is a derivation of the $E_8$ flavor symmetry, 
but it does at least indicate that the $E_8$ symmetry needs to arise automatically.

\subsection{World-sheet structure of strings of  minimal 6d \Nequals{(1,0)} theories}
Finally, let us consider strings of minimal 6d \Nequals{(1,0)} theories,
whose construction in F-theory was recalled in Sec.~\ref{sec:introduction}.
We will continue to use the same symbols there.
Using the 6d anomaly polynomial computed in \cite{Ohmori:2014kda}, we see that
  a charge $Q$ string has the anomaly polynomial  \begin{multline}
I_4(n,Q) = \frac{nQ^2 - (n-2)Q}{2} c_2(L) - \frac{nQ^2 + (n-2)Q}{2} c_2(R) \\
+ \frac{nQ}4 \Tr F^2_G + \frac{(n-2)Q}4 p_1(T) + Q h^\vee_G c_2(I)\label{eq:minimalformula}
\end{multline} by applying our main formula \eqref{eq:concretemainformula}.

We know that an instanton configuration of $\mathfrak{g}$ vector multiplet is charged under the tensor field, such that the instanton number is identified with the charge $Q$ of the instanton-string. This means that the 2d world-sheet theory with \Nequals{(0,4)} supersymmetry of the strings of the minimal 6d \Nequals{(1,0)} theory should at least have a Higgs branch which is the instanton moduli space of gauge group $\mathfrak{g}$ of instanton number $Q$.
This has a quaternionic dimension $h^\vee Q$.

For $n=4$ and $\mathfrak{g}=\mathfrak{so}(8)$, the worldsheet gauge theory is known and is precisely the ADHM construction of charge-$Q$ $\mathrm{so}(8)$ instanton. 
This suggests that, even for other $n$ in Table~\ref{table:foo}, the worldsheet theory is the \Nequals{(0,4)} sigma model on the charge-$Q$ instanton moduli space of gauge algebra $\mathfrak{g}$.

Now, moving along the Higgs branch does not break $\SU(2)_I$ symmetry and the diffeomorphism symmetry of the worldsheet. 
Therefore the terms proportional to $p_1(T)$ and $c_2(I)$ can be obtained straightforwardly at the generic point on the Higgs branch.
There, the string is in fact a finite-sized instanton-string, and we know that there are $4h^\vee Q$ bosonic zero modes and the same number of chiral fermionic zero modes.
From this, we see that the anomaly polynomial should contain the terms \begin{equation}
Qh^\vee_G (\frac{1}{12}p_1(T)+c_2(I)).
\end{equation}
Comparing with \eqref{eq:minimalformula}, we see that we need to have \begin{equation}
h^\vee=3(n-2).
\end{equation}
This explains the curious numerology \eqref{eq:curious} pointed out in Sec.~\ref{sec:introduction}.

\section*{Acknowledgments}
The authors thank Noriaki Watanabe for helpful discussions, and Sunil Mukhi for discussions concerning footnote \ref{footnote:deligne}.
 HS is partially supported by the Programs for Leading Graduate Schools, MEXT, Japan,
via the  Leading Graduate Course for Frontiers of Mathematical Sciences and Physics.
HS is also supported by JSPS Research Fellowship for Young Scientists.
YT is  supported in part by JSPS Grant-in-Aid for Scientific Research No. 25870159,
and in part by WPI Initiative, MEXT, Japan at IPMU, the University of Tokyo.

\bibliographystyle{ytphys}
\small\baselineskip=.9\baselineskip
\bibliography{ref}

\providecommand{\href}[2]{#2}\begingroup\raggedright\begin{thebibliography}{10}

\bibitem{Heckman:2013pva}
J.~J. Heckman, D.~R. Morrison, and C.~Vafa, ``{On the Classification of 6D
  SCFTs and Generalized ADE Orbifolds},''
\href{http://arxiv.org/abs/1312.5746}{{\ttfamily arXiv:1312.5746 [hep-th]}}.

\bibitem{DelZotto:2014hpa}
M.~Del~Zotto, J.~J. Heckman, A.~Tomasiello, and C.~Vafa, ``{6d Conformal
  Matter},'' \href{http://dx.doi.org/10.1007/JHEP02(2015)054}{{\em JHEP}
  {\bfseries 02} (2015) 054},
\href{http://arxiv.org/abs/1407.6359}{{\ttfamily arXiv:1407.6359 [hep-th]}}.

\bibitem{Heckman:2015bfa}
J.~J. Heckman, D.~R. Morrison, T.~Rudelius, and C.~Vafa, ``{Atomic
  Classification of 6D SCFTs},''
\href{http://arxiv.org/abs/1502.05405}{{\ttfamily arXiv:1502.05405 [hep-th]}}.

\bibitem{Bhardwaj:2015xxa}
L.~Bhardwaj, ``{Classification of 6D $\mathcal{N}=(1,0)$ Gauge Theories},''
\href{http://arxiv.org/abs/1502.06594}{{\ttfamily arXiv:1502.06594 [hep-th]}}.

\bibitem{Cordova:2015fha}
C.~Cordova, T.~T. Dumitrescu, and K.~Intriligator, ``{Anomalies,
  Renormalization Group Flows, and the $a$-Theorem in Six-Dimensional (1,0)
  Theories},''
\href{http://arxiv.org/abs/1506.03807}{{\ttfamily arXiv:1506.03807 [hep-th]}}.

\bibitem{Heckman:2015axa}
J.~J. Heckman and T.~Rudelius, ``{Evidence for $c$-Theorems in 6D SCFTs},''
  \href{http://dx.doi.org/10.1007/JHEP09(2015)218}{{\em JHEP} {\bfseries 09}
  (2015) 218},
\href{http://arxiv.org/abs/1506.06753}{{\ttfamily arXiv:1506.06753 [hep-th]}}.

\bibitem{Ohmori:2014kda}
K.~Ohmori, H.~Shimizu, Y.~Tachikawa, and K.~Yonekura, ``{Anomaly Polynomial of
  General 6D SCFTs},'' \href{http://dx.doi.org/10.1093/ptep/ptu140}{{\em PTEP}
  {\bfseries 2014} no.~10, (2014) 103B07},
\href{http://arxiv.org/abs/1408.5572}{{\ttfamily arXiv:1408.5572 [hep-th]}}.

\bibitem{Intriligator:2014eaa}
K.~Intriligator, ``{6D, $ \mathcal{N}=(1,0) $ Coulomb Branch Anomaly
  Matching},'' \href{http://dx.doi.org/10.1007/JHEP10(2014)162}{{\em JHEP}
  {\bfseries 1410} (2014) 162},
\href{http://arxiv.org/abs/1408.6745}{{\ttfamily arXiv:1408.6745 [hep-th]}}.

\bibitem{Haghighat:2013gba}
B.~Haghighat, A.~Iqbal, C.~{Koz\c{}az}, G.~Lockhart, and C.~Vafa,
  ``{M-Strings},'' \href{http://dx.doi.org/10.1007/s00220-014-2139-1}{{\em
  Commun. Math. Phys.} {\bfseries 334} no.~2, (2015) 779--842},
\href{http://arxiv.org/abs/1305.6322}{{\ttfamily arXiv:1305.6322 [hep-th]}}.

\bibitem{Haghighat:2013tka}
B.~Haghighat, C.~Kozcaz, G.~Lockhart, and C.~Vafa, ``{Orbifolds of
  M-strings},'' \href{http://dx.doi.org/10.1103/PhysRevD.89.046003}{{\em Phys.
  Rev.} {\bfseries D89} no.~4, (2014) 046003},
\href{http://arxiv.org/abs/1310.1185}{{\ttfamily arXiv:1310.1185 [hep-th]}}.

\bibitem{Haghighat:2014pva}
B.~Haghighat, G.~Lockhart, and C.~Vafa, ``{Fusing E-strings to heterotic
  strings: $E+E\to H$},''
  \href{http://dx.doi.org/10.1103/PhysRevD.90.126012}{{\em Phys. Rev.}
  {\bfseries D90} no.~12, (2014) 126012},
\href{http://arxiv.org/abs/1406.0850}{{\ttfamily arXiv:1406.0850 [hep-th]}}.

\bibitem{Kim:2014dza}
J.~Kim, S.~Kim, K.~Lee, J.~Park, and C.~Vafa, ``{Elliptic Genus of
  E-Strings},''
\href{http://arxiv.org/abs/1411.2324}{{\ttfamily arXiv:1411.2324 [hep-th]}}.

\bibitem{Haghighat:2014vxa}
B.~Haghighat, A.~Klemm, G.~Lockhart, and C.~Vafa, ``{Strings of Minimal 6D
  SCFTs},'' \href{http://dx.doi.org/10.1002/prop.201500014}{{\em Fortsch.
  Phys.} {\bfseries 63} (2015) 294--322},
\href{http://arxiv.org/abs/1412.3152}{{\ttfamily arXiv:1412.3152 [hep-th]}}.

\bibitem{Hohenegger:2015cba}
S.~Hohenegger, A.~Iqbal, and S.-J. Rey, ``{M-Strings, Monopole Strings, and
  Modular Forms},'' \href{http://dx.doi.org/10.1103/PhysRevD.92.066005}{{\em
  Phys. Rev.} {\bfseries D92} no.~6, (2015) 066005},
\href{http://arxiv.org/abs/1503.06983}{{\ttfamily arXiv:1503.06983 [hep-th]}}.

\bibitem{Gadde:2015tra}
A.~Gadde, B.~Haghighat, J.~Kim, S.~Kim, G.~Lockhart, and C.~Vafa, ``{6D String
  Chains},''
\href{http://arxiv.org/abs/1504.04614}{{\ttfamily arXiv:1504.04614 [hep-th]}}.

\bibitem{Kim:2015fxa}
J.~Kim, S.~Kim, and K.~Lee, ``{Higgsing Towards E-Strings},''
\href{http://arxiv.org/abs/1510.03128}{{\ttfamily arXiv:1510.03128 [hep-th]}}.

\bibitem{Haghighat:2016jjf}
B.~Haghighat and W.~Yan, ``{M-Strings in Thermodynamic Limit: Seiberg-Witten
  Geometry},''
\href{http://arxiv.org/abs/1607.07873}{{\ttfamily arXiv:1607.07873 [hep-th]}}.

\bibitem{Morrison:2012np}
D.~R. Morrison and W.~Taylor, ``{Classifying Bases for 6D F-Theory Models},''
  \href{http://dx.doi.org/10.2478/s11534-012-0065-4}{{\em Central Eur. J.
  Phys.} {\bfseries 10} (2012) 1072--1088},
\href{http://arxiv.org/abs/1201.1943}{{\ttfamily arXiv:1201.1943 [hep-th]}}.

\bibitem{Beem:2013sza}
C.~Beem, M.~Lemos, P.~Liendo, W.~Peelaers, L.~Rastelli, and B.~C. van Rees,
  ``{Infinite Chiral Symmetry in Four Dimensions},''
  \href{http://dx.doi.org/10.1007/s00220-014-2272-x}{{\em Commun. Math. Phys.}
  {\bfseries 336} no.~3, (2015) 1359--1433},
\href{http://arxiv.org/abs/1312.5344}{{\ttfamily arXiv:1312.5344 [hep-th]}}.

\bibitem{Lemos:2015orc}
M.~Lemos and P.~Liendo, ``{$\mathcal{N}=2$ Central Charge Bounds from $2d$
  Chiral Algebras},'' \href{http://dx.doi.org/10.1007/JHEP04(2016)004}{{\em
  JHEP} {\bfseries 04} (2016) 004},
\href{http://arxiv.org/abs/1511.07449}{{\ttfamily arXiv:1511.07449 [hep-th]}}.

\bibitem{Deligne1}
P.~Deligne, ``{La s\'erie exceptionnelle de groupes de Lie},'' {\em {C. R.
  Acad. Sci. Paris}} {\bfseries 322} (1996) 321--326.

\bibitem{Deligne2}
P.~Deligne and R.~de~Man, ``{La s\'erie exceptionnelle de groupes de Lie II},''
  {\em {C. R. Acad. Sci. Paris}} {\bfseries 323} (1996) 577--582.

\bibitem{Grassi:2000we}
A.~Grassi and D.~R. Morrison, ``{Group Representations and the Euler
  Characteristic of Elliptically Fibered Calabi-Yau Threefolds},''
\href{http://arxiv.org/abs/math/0005196}{{\ttfamily arXiv:math/0005196
  [math.AG]}}.

\bibitem{Mathur:1988na}
S.~D. Mathur, S.~Mukhi, and A.~Sen, ``{On the Classification of Rational
  Conformal Field Theories},''
\href{http://dx.doi.org/10.1016/0370-2693(88)91765-0}{{\em Phys. Lett.}
  {\bfseries B213} (1988) 303--308}.

\bibitem{Cvitanovic}
P.~Cvitanovi{\'c}, \href{http://dx.doi.org/10.1515/9781400837670}{{\em Group
  theory}}.
\newblock Princeton University Press, Princeton, NJ, 2008.
\newblock \url{http://birdtracks.eu/version9.0/index.html}.

\bibitem{Kim:2016foj}
H.-C. Kim, S.~Kim, and J.~Park, ``{6D Strings from New Chiral Gauge
  Theories},''
\href{http://arxiv.org/abs/1608.03919}{{\ttfamily arXiv:1608.03919 [hep-th]}}.

\bibitem{Sagnotti:1992qw}
A.~Sagnotti, ``{A Note on the Green-Schwarz mechanism in open string
  theories},'' \href{http://dx.doi.org/10.1016/0370-2693(92)90682-T}{{\em Phys.
  Lett.} {\bfseries B294} (1992) 196--203},
\href{http://arxiv.org/abs/hep-th/9210127}{{\ttfamily arXiv:hep-th/9210127}}.

\bibitem{Sadov:1996zm}
V.~Sadov, ``{Generalized Green-Schwarz Mechanism in F Theory},''
  \href{http://dx.doi.org/10.1016/0370-2693(96)01134-3}{{\em Phys.Lett.}
  {\bfseries B388} (1996) 45--50},
\href{http://arxiv.org/abs/hep-th/9606008}{{\ttfamily arXiv:hep-th/9606008}}.

\bibitem{Freed:1998tg}
D.~Freed, J.~A. Harvey, R.~Minasian, and G.~W. Moore, ``{Gravitational Anomaly
  Cancellation for M-theory Fivebranes},'' {\em Adv. Theor. Math. Phys.}
  {\bfseries 2} (1998) 601--618,
\href{http://arxiv.org/abs/hep-th/9803205}{{\ttfamily arXiv:hep-th/9803205}}.

\bibitem{Henningson:2004dh}
M.~Henningson, ``{Self-Dual Strings in Six Dimensions: Anomalies, the
  ADE-Classification, and the World-Sheet WZW-Model},''
  \href{http://dx.doi.org/10.1007/s00220-005-1324-7}{{\em Commun. Math. Phys.}
  {\bfseries 257} (2005) 291--302},
\href{http://arxiv.org/abs/hep-th/0405056}{{\ttfamily arXiv:hep-th/0405056}}.

\bibitem{Berman:2004ew}
D.~S. Berman and J.~A. Harvey, ``{The Self-Dual String and Anomalies in the
  M5-Brane},'' \href{http://dx.doi.org/10.1088/1126-6708/2004/11/015}{{\em
  JHEP} {\bfseries 0411} (2004) 015},
\href{http://arxiv.org/abs/hep-th/0408198}{{\ttfamily arXiv:hep-th/0408198}}.

\bibitem{Henningson:2005hd}
M.~Henningson and E.~P.~G. Johansson, ``{Dyonic Anomalies},''
  \href{http://dx.doi.org/10.1016/j.physletb.2005.09.011}{{\em Phys. Lett.}
  {\bfseries B627} (2005) 203--207},
\href{http://arxiv.org/abs/hep-th/0508103}{{\ttfamily arXiv:hep-th/0508103}}.

\bibitem{Kim:2012wc}
H.~Kim and P.~Yi, ``{D-Brane Anomaly Inflow Revisited},''
  \href{http://dx.doi.org/10.1007/JHEP02(2012)012}{{\em JHEP} {\bfseries 02}
  (2012) 012},
\href{http://arxiv.org/abs/1201.0762}{{\ttfamily arXiv:1201.0762 [hep-th]}}.

\bibitem{Tong:2014yna}
D.~Tong, ``{The Holographic Dual of $AdS_{3} \times S^{3} \times S^{3} \times
  S^{1}$},'' \href{http://dx.doi.org/10.1007/JHEP04(2014)193}{{\em JHEP}
  {\bfseries 04} (2014) 193},
\href{http://arxiv.org/abs/1402.5135}{{\ttfamily arXiv:1402.5135 [hep-th]}}.

\bibitem{Cordova:2015vwa}
C.~Cordova, T.~T. Dumitrescu, and X.~Yin, ``{Higher Derivative Terms, Toroidal
  Compactification, and Weyl Anomalies in Six-Dimensional (2,0) Theories},''
\href{http://arxiv.org/abs/1505.03850}{{\ttfamily arXiv:1505.03850 [hep-th]}}.

\end{thebibliography}\endgroup

\end{document}